\documentclass[aps, pra, twocolumn, 10pt, superscriptaddress, showkeys, longbibliography]{revtex4-1}
\usepackage{color}
\usepackage{graphicx}
\usepackage{amsfonts}
\usepackage{amsmath}
\usepackage{amssymb}
\usepackage{braket}
\usepackage{mathtools}
\usepackage{siunitx}
\usepackage{enumerate}
\usepackage[pdftex,
pdfauthor={Matteo Mariantoni},
pdftitle={High-Density Qubit Wiring: Pin-Chip Bonding for Fully Vertical Interconnects},
pdfkeywords={Quantum Computing; Scalable Qubit Architectures; High-Density Wiring; Pin-Chip Bonding; Fully Vertical Interconnects; Rectangular Coax Ribbon Cables; Superconducting Qubits; Large-Scale Quantum Computers}]{hyperref}

\begin{document}

\title{High-Density Qubit Wiring: Pin-Chip Bonding for Fully Vertical Interconnects}

\author{M.~Mariantoni}
\email[Corresponding author: ]{matteo.mariantoni@uwaterloo.ca}
\affiliation{Institute for Quantum Computing, University of Waterloo, 200 University Avenue West, Waterloo, Ontario N2L 3G1, Canada}
\affiliation{Department of Physics and Astronomy, University of Waterloo, 200 
University Avenue West, Waterloo, Ontario N2L 3G1, Canada}

\author{A.V.~Bardysheva}

\date{\today}

\begin{abstract}
Large-scale quantum computers with more than~$10^5$ qubits will likely be built within the next decade. Trapped ions, semiconductor devices, and superconducting qubits among other physical implementations are still confined in the realm of medium-scale quantum integration ($\sim 100$ qubits); however, they show promise toward large-scale quantum integration. Building large-scale quantum processing units will require truly scalable control and measurement classical coprocessors as well as suitable wiring methods. In this blue paper, we introduce a fully vertical interconnect that will make it possible to address~$\sim 10^5$ superconducting qubits fabricated on a single silicon or sapphire chip: \textit{Pin-chip bonding.} This method permits signal transmission from~DC to~$\sim \SI{10}{\giga\hertz}$, both at room temperature and at cryogenic temperatures down to~$\sim \SI{10}{\milli\kelvin}$. At temperatures below~$\sim \SI{1}{\kelvin}$, the on-chip wiring contact resistance is close to zero and all signal lines are in the superconducting state. High-density wiring is achieved by means of a fully vertical interconnect that interfaces the qubit array with a network of rectangular coaxial ribbon cables. Pin-chip bonding is fully compatible with classical high-density test board applications as well as with other qubit implementations.
\end{abstract}


\maketitle

\section{INTRODUCTION}
	\label{sec:INTRODUCTION}

Practical quantum computers are closer to reality than ever before~\cite{Neill:2018}. Among a variety of physical implementations~\cite{Ladd:2010}, those based on superconducting quantum circuits~\cite{Clarke:2008, Wendin:2017} show promise for the realization of medium-scale quantum computers comprising on the order of~$100$ qubits~\cite{Knight:2017, Giles:2018}. It is expected that these computers will allow us to tackle problems with real-world applications, such as the quantum simulation of chemical reactions, quantum-assisted optimization, and quantum sampling~\cite{Montanaro:2016}. Most importantly, medium-scale quantum computers will serve as a test bed for quantum error correction~(QEC) algorithms~\cite{Gottesman:2010}. These algorithms will make it possible to build a fault-tolerant digital quantum computer and, thus, to harness the full power of quantum algorithms~\cite{Nielsen:2000}. Building a useful, QEC-based quantum computer will require large-scale quantum integration~(LSQI), where quantum processing units~(QPUs) are comprised of two-dimensional qubit arrays with~$\sim 10^5$ qubits.

Superconducting quantum circuits are planar devices fabricated from aluminum~(Al) thin films deposited on silicon~(Si) substrates and patterned by means of photolithography techniques similar to those used in classical integrated-circuit technology. The Al~films are patterned to realize capacitive and inductive elements, $C$ and $L$, respectively. These elements can be used to implement harmonic oscillators, i.e., linear resonators. A nonlinear inductor can be realized by means of a Josephson tunnel junction, which comprises a pair of Al~islands separated by an ultra-thin insulating layer of Al~oxide. A suitable combination of linear and nonlinear circuit elements allows us to implement anharmonic oscillators, i.e., nonlinear resonators. Linear and nonlinear superconducting resonators are designed to have a resonance frequency~$f_0$ in the microwave range, typically~$f_0 \in [ 5, 10] \,\SI{}{\giga\hertz}$. Thus, they can be controlled with signals similar to those used in mobile telephones. Circuit operation at~DC is also required in certain designs.

Microwave resonators can be prepared in the quantum regime by cooling them to a temperature~$T \sim \SI{10}{\milli\kelvin}$ in a dilution refrigerator~(DR) and limiting their population to the lowest quantum-mechanical states~$\{ \ket{0}, \ket{1}, \ket{2}, \ldots \}$. Low temperatures are required to reduce superconductor excitations (quasiparticles) and thermal noise. In the case of nonlinear resonators, it is possible to isolate states~$\ket{0}$ and $\ket{1}$ from all higher states, thus forming a qubit. One of the most widely used superconducting qubits is the transmon qubit~\cite{Koch:2007}, where an~$\sim \SI{100}{\micro\meter\squared}$ Al~island is connected to a pair of~$\sim \SI{100}{\nano\meter\squared}$ Josephson tunnel junctions. To date, ``Xmon-type'' transmon qubits~\cite{Barends:2013} have been used to realize a~QPU comprising a one-dimensional array of nine qubits integrated on a single chip~\cite{Kelly:2015}. Moreover, transmon qubits are at the core of the medium-scale quantum computers that are being developed by IBM Q, Google Santa Barbara, Intel-Delft, and Rigetti~\cite{Knight:2017, Giles:2018, Versluis:2017, Otterbach:2017}.

The two main challenges to move from medium- to large-scale quantum computers are low error rates and true scalability~\cite{Martinis:2015}. These two objectives must be developed simultaneously, as progressively scalable control and measurement classical coprocessors and qubit wiring methods have to be compatible with high-fidelity qubit operations (e.g., one- and two-qubit gates and qubit measurement). In this blue paper, we introduce a fully vertical interconnect that will make it possible to address~$\sim 10^5$ superconducting qubits fabricated on a single~Si or sapphire chip: \textit{Pin-chip bonding.} This method permits signal transmission from~DC to $\sim \SI{10}{\giga\hertz}$, both at room temperature and at cryogenic temperatures down to~$\sim \SI{10}{\milli\kelvin}$. At temperatures below~$\sim \SI{1}{\kelvin}$, the on-chip wiring contact resistance is close to zero and all signal lines are in the superconducting state. High-density wiring is achieved by means of a fully vertical interconnect that interfaces the qubit array with a network of rectangular coaxial ribbon cables~\cite{Tuckerman:2016}. Pin-chip bonding is fully compatible with classical high-density test board applications as well as with other qubit implementations. Here, we focus on superconducting qubit applications as they set very stringent specifications, therefore representing an excellent benchmarking platform.

\section{QUBIT WIRING: STATE OF THE ART AND CHALLENGES AHEAD}
	\label{sec:QUBIT:WIRING::STATE:OF:THE:ART:AND:CHALLENGES:AHEAD}

Arguably, the first method that made it possible to wire up superconducting QPUs with~$\sim 100$ qubits has been the quantum socket~\cite{Bejanin:2016}. Traditional wiring methods based on wire bonds~\cite{Wenner:2011:a} allow accessing the~QPU only laterally, through the four edges of the QPU chip. The quantum socket is based on a vertical wiring approach and, thus, permits us to reach any area on the chip surface. The vertical wires are realized as spring-loaded micro coaxes, i.e., custom-made Pogo pins, which can carry signals in the microwave range as well as DC with low loss. The quantum socket has been used routinely to measure high-quality planar superconducting resonators~\cite{Earnest:2018} as well as qubits~\cite{Bronn:2018}.

Vertical connectivity is one of the main innovations introduced by the quantum socket. However, the dimensions of the Pogo pins are too large to extend the socket to large-scale quantum computers. In fact, fabricating even smaller Pogo pins that those used in the works of Refs.~\cite{Bejanin:2016, Bronn:2018} is challenging. The presence of the springs in the Pogo pins, in particular, makes further miniaturization exceedingly hard. Due to this limitation, the footprint of one fully coaxial Pogo pin assembly with characteristic impedance~$Z_{\textrm{c}} \approx \SI{50}{\ohm}$ is~$\approx \SI{1}{\milli\meter}$. This dimension can be reduced to~$\approx \SI{0.5}{\milli\meter}$ by making an assembly with~$Z_{\textrm{c}} \approx \SI{25}{\ohm}$. Considering the largest Xmon transmon qubit is~$\sim \SI{500}{\micro\meter} \times \SI{500}{\micro\meter}$~\cite{Kelly:2015}, the Pogo pin approach will not allow for high-dense wiring as the smallest fully coaxial Pogo pin assembly is larger than the footprint of one (large) qubit. Additionally, the Pogo pin approach in the work of Ref.~\cite{Bronn:2018} interfaces a Pogo-pin array and the qubit control and measurement network by manes of a printed circuit board connected to a high-throughput commercial connector. The work in Ref.~\cite{Bejanin:2016}, instead, interfaces the Pogo pins with~EZ~$47$ coaxial cables. Both approaches effectively increase significantly the wiring footprint when interfacing the QPU with the qubit control and measurement network and, thus, are not suitable for a large-scale quantum computer.

In an effort to develop more scalable qubit wiring methods, MIT Lincoln Laboratory and Google Santa Barbara have built multi-chip QPUs where flip-chip technology is used to bond a chip with qubits (quantum chip) to a chip with control and measurement circuitry (classical chip)~\cite{Rosenberg:2017, Foxen:2017}. A typical bonding procedure consists of:
\begin{enumerate}[(1)]
	\item Patterning a two-dimensional array of indium~(In) bumps on the top surface of the classical chip, with mating In~pads on the bottom surface of the quantum chip;
	\item flipping the quantum chip over the classical chip and aligning the two chips;
	\item compressing the two chips with a pressure~$p \in ( 10, 20 ) \SI{}{\newton\per\milli\meter\squared}$.
\end{enumerate}
Indium bumps are pillars up to~\SI{30}{\micro\meter} tall with pre- and post-compression diameters~$d = \SI{15}{\micro\meter}$ and $d = \SI{30}{\micro\meter}$, respectively. Flip-technology makes it possible to access qubits in two-dimensional arrays, an otherwise impossible task using only wire bonds. However, it still relies on wire bonds that are ultimately used to interface the four edges of the classical chip with the control network. Wire bonds are used even in the most recent proposal on qubit wiring and integration by MIT Lincoln Laboratory~\cite{Das:2018}.

Given~$n_{\textrm{w}}$ wire bonds per chip edge, the total number of wires available with flip chip is~$N_{\textrm{w}} = 4 n_{\textrm{w}}$. On the other hand, the total number of qubits in a two-dimensional array is~$N_{\textrm{q}} = n^2_{\textrm{q}}$, where~$n_{\textrm{q}}$ is the number of qubits per chip edge. The linear scaling of the wires fails to match the quadratic scaling of the qubits, therefore limiting flip chip to medium-scale QPUs.

Present implementations of flip chip can potentially be extended to large-scale QPUs with~$\sim 10^5$ qubits if it were possible to include one controller per qubit on the classical chip (similar arguments would apply to a qubit measurer), fulfilling three conditions~\cite{McDermott:2018}:
\begin{enumerate}[(1)]
	\item Qubit and controller have similar physical footprints;
	\item each controller dissipates a power~$P_{\textrm{c}} \sim \SI{1}{\nano\watt}$;
	\item controllers are heavily demultiplexed~(DEMUXed).
\end{enumerate}
Two types of cryogenic controllers have been proposed, single flux quantum~(SFQ)~\cite{McDermott:2018} and cryo-CMOS~\cite{Patra:2018}. McDermott~\textit{et al.} have shown that for the former~$P_{\textrm{c}} \sim \SI{100}{\nano\watt}$ and for the latter~$P_{\textrm{c}} \sim \SI{10}{\micro\watt}$. Hence, both proposals suggest to operate the controller system at the~\SI{3}{\kelvin} stage of a pulse tube cooler, which has a cooling power of~$\sim \SI{1}{\watt}$ but is~$\sim \SI{1}{\meter}$ away from the quantum chip.

This approach thus hinders the use of flip chip, unless it were possible to devise a multi-chip QPU where the quantum chip is operated at~$\sim \SI{10}{\milli\kelvin}$, the classical chip (with controllers) at~\SI{3}{\kelvin}, and the two chips are separated by an interposer chip for heat shielding. Assuming~\SI{100}{\percent} heat shielding from the interposer substrate, metalized through-silicon vias~(TSVs) will be needed to connect the classical and quantum chip vertically. Using niobium-titanium~(Nb-Ti) for the vias, the heat load on the quantum chip would be~$\sim \SI{20}{\milli\watt}$~\cite{McDermott:2018}, which is likely too high even for a specially designed~DR.

Lastly, we could elect to operate the entire QPU at~\SI{3}{\kelvin} and refrigerate the quantum chip to~$\sim \SI{10}{\milli\kelvin}$ by means of some sort of resolved-sideband cooling (microwave-induced cooling)~\cite{Valenzuela:2006}, possibly followed by algorithmic cooling~\cite{Park:2016}. However, this approach will result in a significant runtime overhead in the quantum computation as well as, likely, one- and two-qubit gates with higher error rates. Therefore, a fully vertical wiring method with footprint smaller than the qubit footprint and uniform throughout the entire control and measurement network is highly appealing.

In this work, we show that it is possible to fabricate a fully coaxial pin-chip assembly with a footprint of~\SI{0.4}{\milli\meter} and $Z_{\textrm{c}} \approx \SI{50}{\ohm}$ or \SI{0.2}{\milli\meter} and $Z_{\textrm{c}} \approx \SI{25}{\ohm}$. Additionally, we show that such a footprint can be maintained throughout the entire control and measurement network, if pin-chip is suitably interfaced with a network of rectangular coaxial ribbon cables.

\section{PIN-CHIP BONDING FOR FULLY VERTICAL INTERCONNECTS}
	\label{sec:PIN-CHIP:BONDING:FOR:FULLY:VERTICAL:INTERCONNECTS}

\begin{figure*}
\includegraphics[scale=0.275]{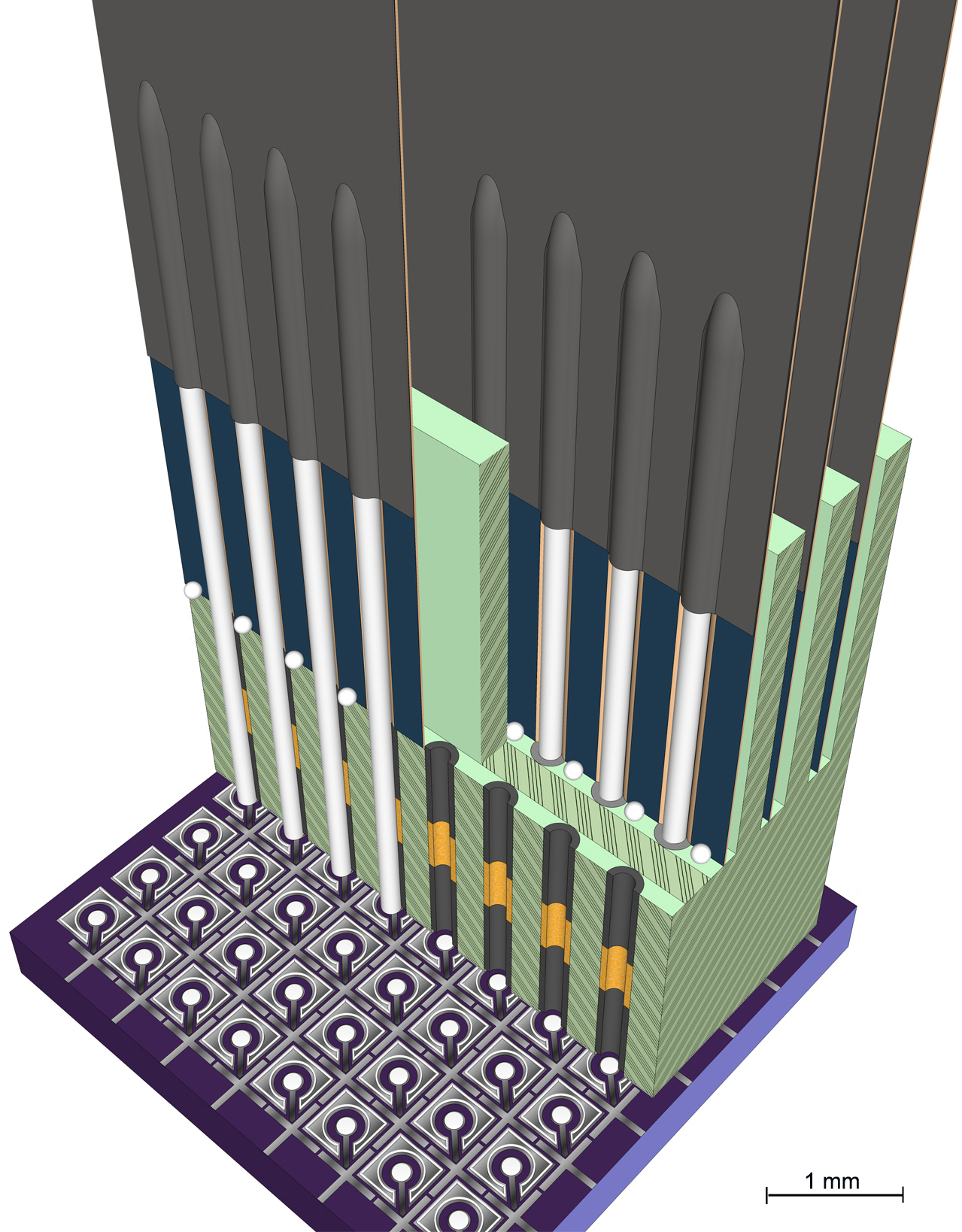}
	\caption{Overall view of pin-chip bonding for fully vertical interconnects, with cross-section of the interposer block showing the channels and through holes where the ribbon cables are inserted. Salmon: Polyimide tape. Dark gray: Nb~films. The PTFE spacers (dark gray) sandwich a STYCAST block (yellow). Pins are white. On the quantum chip: Violet is~Si; middle-metallic gray is~Al; white is~In~(curbs).}
\label{figure:1:mariantoni}
\end{figure*}

\begin{figure*}
\includegraphics[scale=0.35]{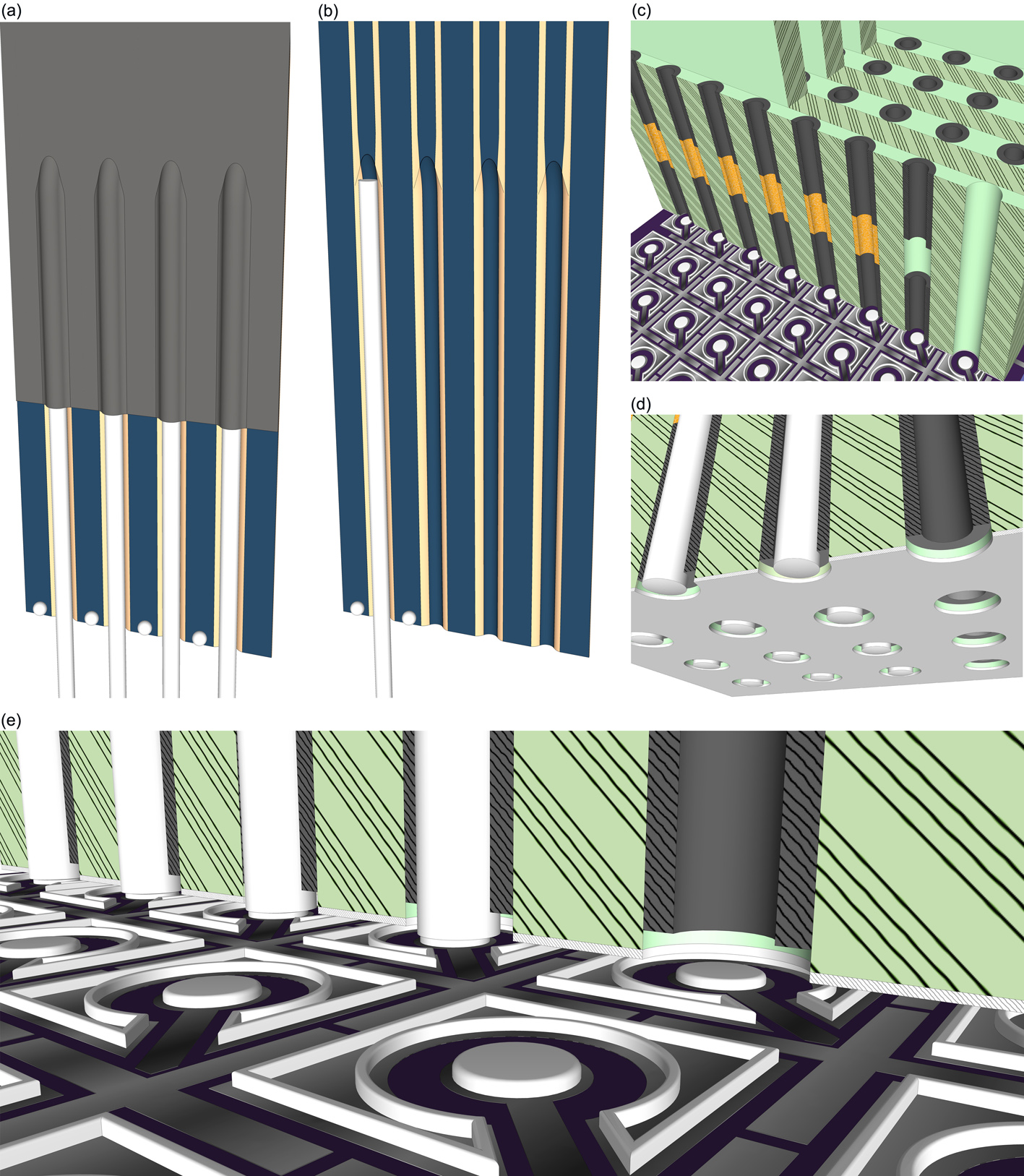}
	\caption{The three main elements of pin-chip bonding. (a) One ribbon cable forming a cable-pin-cable assembly. A row of In~solder balls is shown. (b) Same as~(a), but without the front cable, exposing pins, polyimide tape, and metallic film; only the first pin is shown. The CPW tapering at the top of each pin is visible. (c) Cross-section of the interposer block: Top view. (d) Cross-section of the interposer block: Bottom view. (e) Pin-chip bonding on the quantum chip. The color coding is as in Fig.~\ref{figure:1:mariantoni}.}
\label{figure:2a:2e:mariantoni}
\end{figure*}

\begin{figure}
\includegraphics[scale=0.4]{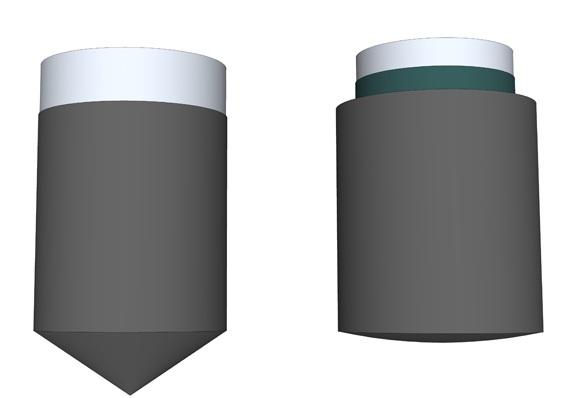}
	\caption{Conical (left) and spherical (right) bonding. See main text for details.}
\label{figure:3:mariantoni}
\end{figure}

In order to better understand the limitations of flip chip and the advantages of pin-chip bonding, we need to first outline the fundamental conditions for dense qubit wiring. As an example, consider a square qubit array, where each qubit has a footprint~$\ell^2_{\textrm{q}} = \SI{500}{\micro\meter} \times \SI{500}{\micro\meter}$, as in the work of Ref.~\cite{Kelly:2015}\footnote{Note that, pin-chip bonding is not limited to this specific qubit footprint. Smaller qubit footprints and different qubit designs can me considered as well.}, with a qubit pitch~$p_{\textrm{q}}$ equal to the qubit lateral dimension~$\ell_{\textrm{q}}$, $p_{\textrm{q}} = \ell_{\textrm{q}}$. Further consider a square wire array with wire pitch~$p_{\textrm{w}}$. In the case of flip chip, the minimum~$p_{\textrm{w}}$ is set by the smallest available~Al or gold~(Au) bonding wire with diameter~$d = \SI{18}{\micro\meter}$~\footnote{See, e.g., \url{http://www.scanditron.com/sites/default/files/material/heraeus_bondingwire_brochure.pdf} or \url{https://www.cirexx.com/wire-bonding/}.}. Since the qubit control signals are in the microwave regime, they must be carried by suitable transmission lines. Thus, it is necessary to use at least three bonding wires (two grounds and one signal) next to each other to realize such a line. Assuming an optimistic spacing of~\SI{10}{\micro\meter} between adjacent wires and sharing grounds between two adjacent transmission lines, we obtain~$p_{\textrm{w}} = \SI{56}{\micro\meter}$.

The \textit{first condition} for dense qubit wiring is
\begin{equation}
\frac{p_{\textrm{w}}}{p_{\textrm{q}}} \leqslant 1 \, ,
	\label{eq:dense:wiring:condition:1}
\end{equation}
i.e., the overall wire lateral dimension cannot be larger than the qubit dimension. This condition can easily be fulfilled in the case of flip chip. The \textit{second condition} applies to square wire arrays with \textit{lateral access}, as flip chip, imposing a constraint on the lateral dimension of the quantum chip, $\ell$,
\begin{equation}
N_{\textrm{q}} \equiv \left( \frac{\ell}{p_{\textrm{q}}} \right)^2 = 4 \frac{\ell}{p_{\textrm{w}}} \equiv N_{\textrm{w}} \, .
	\label{eq:dense:wiring:condition:2}
\end{equation}
In Eq.~(\ref{eq:dense:wiring:condition:2}) we assume only one wire per qubit (i.e., a significant DEMUX). The nontrivial solution to Eq.~(\ref{eq:dense:wiring:condition:2}),
\begin{equation}
\ell = 4 \frac{p^2_{\textrm{q}}}{p_{\textrm{w}}} \, ,
	\label{eq:solution:condition:2}
\end{equation}
is the intercept between the linear scaling curve of the wires and the quadratic scaling curve of the qubits. Notably, this condition does not prescribe a practical method to reach any qubit in the array, providing instead an upper bound for~$N_{\textrm{q}}$. For an optimistic implementation of flip chip with~$p_{\textrm{q}} = \SI{500}{\micro\meter}$ and $p_{\textrm{w}} = \SI{56}{\micro\meter}$, we obtain~$\ell \approx \SI{18}{\milli\meter}$ that corresponds to~$N_{\textrm{q}} = 1296$. This is the maximum number of qubits that makes sense to address using flip chip. For any given larger chip there would not be sufficient wires on the chip edges to address all qubits on a two-dimensional array. Flip chip is thus confined to the realm of medium-scale QPUs.

Figure~\ref{figure:1:mariantoni} shows a schematic representation of the qubit wiring method proposed here, pin-chip bonding for fully vertical interconnects, which will make it possible to implement large-scale QPUs. This method is based on rectangular coaxial ribbon cables attached to micro-sized pins, forming coaxial transmission lines by means of an interposer block; the tip of each pin is bonded vertically to mating pads on the quantum chip using~In. The quantum chip is patterned with a square array of~$[ 5, 30 ] \, \SI{}{\micro\meter}$ thick circular pads made from~In with~$d_{\textrm{pad}} \in [ 100, 200 ] \,\SI{}{\micro\meter}$; the In~pads can be fabricated above an underlying Al~thin film, or other films. Each pad is electrically connected to an on-chip~Al (or Nb) trace used to reach a qubit. The pad array is in a one-to-one correspondence with a mating pin array. Each pin is~$\sim [ 15, 25 ] \,\SI{}{\milli\meter}$ long and has a core made from~SUS~304 Austenite stainless steel (or similar) with~$d \in [ 78, 178 ] \,\SI{}{\micro\meter}$, coated first with a~$\sim \SI{1}{\micro\meter}$ thick titanium nitride~(TiN) film and then with a~\SI{10}{\micro\meter} thick (or possibly thicker) In~film resulting in~$d_{\textrm{pin}} = d_{\textrm{pad}}$. The~TiN coating, which can be applied using physical vapor deposition, guarantees that microwave signals are carried within a superconducting region at~$T \sim \SI{10}{\milli\kelvin}$ and acts as an interdiffusion barrier~\cite{Foxen:2017}, if necessary. The In~film can be applied by means of thermal evaporation [other materials such as tin-lead~(Sn-Pb) eutectic alloys may be used] and serves as the bonding agent between pin and pad.

We fabricate rectangular coaxial ribbon cables, each consisting of a~\SI{20}{\micro\meter} thick flexible polyimide tape coated with a~\SI{250}{\nano\meter} thick Nb~film (similar to the work in Ref.~\cite{Tuckerman:2016}). This film is patterned to form a set of parallel transmission lines in a coplanar waveguide~(CPW) design, which are then coated with a~\SI{10}{\micro\meter} thick (or possibly thicker) Sn-Pb~film. The pins are attached to the cables following this procedure:
\begin{enumerate}[(1)]
	\item The~$\lesssim \SI{10}{\milli\meter}$ tail of each pin is placed on one signal trace of a ribbon cable.
	\item another ribbon cable is flipped over and aligned with the underlying cable with a vertical offset of~$\sim \SI{1}{\milli\meter}$, exposing part of its bottom;
	\item the cable-pin-cable assembly is compressed and soldered in a reflow oven at~$T \gtrsim \SI{183}{\degreeCelsius}$, with the front segment of each pin free hanging at the bottom of the assembly [see Fig.~\ref{figure:2a:2e:mariantoni}~(a) and (b)].
\end{enumerate}
After the reflow step, In~solder balls with~$d \lesssim \SI{50}{\micro\meter}$ are pressed on the exposed conductor of the ground traces, each of width~$G = \SI{50}{\micro\meter}$. The In~balls are placed on the ground traces by means of an~XY stage equipped with micro tweezers.

Note that a slight tapering of the transmission line on the ribbon cables must be engineered in order to transition smoothly from the cable region overlapping with the circular pin to the flat cable region. This taring allows us not to increase the wiring footprint [see Fig.~\ref{figure:2a:2e:mariantoni}~(b)].

To form a coaxial transmission line and perform pin-pad alignment we fabricate an interposer block made of, e.g., oxygen-free high thermal conductivity~(OFHC) copper~(Cu), where a square array of through holes each with~$d_{\textrm{hole}} \in [ 200, 300] \,\SI{}{\micro\meter}$ is made by way of fast hole drilling electrical discharge machining. Each row of holes resides at the bottom of a channel of width~$\gtrsim [ 140, 240 ] \,\SI{}{\micro\meter}$ and depth~$\sim \SI{1}{\milli\meter}$ [see Fig.~\ref{figure:1:mariantoni} and Fig.~\ref{figure:2a:2e:mariantoni}~(c)]. The channel~(groove) is fabricated by means of single-point diamond turning~(SPDT)~\cite{Higginbottom:2018}, which provides extremely high nanometric accuracy and precision and allows to trench narrow channels in~Cu and possibly other materials. The aspect ratio of our channels is at a minimum~$\sim \SI{140}{\micro\meter} / \SI{1}{\milli\meter} = 0.14$, which is not terribly hard to achieve with~SPDT. Boss extruders or dowels are used to align the interposer to a sample holder housing the quantum chip.

The exposed part of a cable-pin-cable assembly is inserted in one of the channels, with the front segment of each pin threaded through a matching hole in the interposer. Prior to pin insertion, the holes are filled with, e.g., STYCAST~1266~A/B~\footnote{See, e.g., \url{http://research.physics.illinois.edu/bezryadin/labprotocol/stycast1266.pdf}.}, which has a relative electric permittivity~$\varepsilon_{\textrm{r}} \approx 3$. Two polytetrafluoroethylene~(PTFE) spacers [see Fig.~\ref{figure:2a:2e:mariantoni}~(c) and (d)] are used to separate the pin from the hole's wall, realizing a coaxial transmission line. During insertion the pins are protected by means of a~$\sim \SI{250}{\nano\meter}$ thick photoresist film; the photoresist on the pin's tip is stripped prior to bonding (the resist on the exposed part of the pin simply adds to the coaxial transmission line dielectric). After a curing step at~$T \approx \SI{60}{\degreeCelsius}$, the entire assembly is heated in a vacuum oven at~$T \gtrsim \SI{157}{\degreeCelsius}$, soldering the In~balls to one wall of the channel. A dielectric spray or filling~(grease) is used to isolate the exposed part of the transmission lines from the interposer block. The tip of each pin is adjusted to be flush with the interposer bottom surface [see Fig.~\ref{figure:2a:2e:mariantoni}~(d)].
All tips must be aligned vertically with a tolerance of at least $\mp \SI{2.5}{\micro\meter}$ pushing the pin array onto mesa stops patterned on an auxiliary Si~chip. This chip is eventually substituted with the quantum chip when performing the actual pin-chip bonding step.

Note that many rows of rectangular coaxial ribbon cables will be in close proximity when wiring a large two-dimensional array of qubits. In order to shield adjacent cable rows and avoid large inter-cable crosstalk, the two outer surfaces of each polyimide tape must be coated with a~\SI{250}{\nano\meter} thick Nb~film. This conductive (superconductive below~$T \approx \SI{9.2}{\kelvin}$) film not only acts as an electromagnetic shield between adjacent cable rows, but also effectively results in a stripline transmission line~\cite{Collin:2001}. This helps mitigate significantly the intra-cable crosstalk between adjacent transmission lines on the same tape due to a field confinement between the signal trace and outer ground larger than between the signal trace and lateral CPW grounds.

Depending on the chosen pin diameter, the characteristic impedance of the coaxial transmission line in the interposer is~$Z_{\textrm{c}} \approx [24, 14] \,\SI{}{\ohm}$. We propose to depart from~$Z_{\textrm{c}} = \SI{50}{\ohm}$ currently used for qubit control lines and resort to smaller impedances, where the inner conductor diameter is maximized for mechanical robustness and the outer conductor diameter is reduced to obtain smaller lines. We note that the vast majority of the control electronics used for medium-scale quantum computers is already fully custom made and, thus, can be designed with practically arbitrary values of~$Z_{\textrm{c}}$.

The bonding procedure is realized as follows:
\begin{enumerate}[(1)]
	\item The thin In~oxide layer on the pin's outer surface is etched with hydrochloric~acid~\cite{Brecht:2017} and the In~oxide film on the pad with a plasma etch~\cite{Foxen:2017};
	\begin{enumerate}[(a)]
		\item \textit{Conical bonding} (see Fig.~\ref{figure:3:mariantoni}~left) is realized with a sharp conical pin tip that penetrates into the In~layer of the pad. In this case, the pin outer surface should not be coated with~In and pre-bonding oxide cleaning may be unnecessary. Using a similar pressure per pin as for flip chip, our experience with Pogo pins~\cite{Bejanin:2016} indicates that the pin will puncture only the first~$\lesssim \SI{1}{\micro\meter}$ of a~\SI{10}{\micro\meter} thick In~pad;
		\item Alternatively, \textit{Spherical bonding} (see Fig.~\ref{figure:3:mariantoni}~right) is realized using a rounded spherical pin tip that is compressed onto the underlying pad with a similar pressure as for flip chip, corresponding to~$\approx 280$ to \SI{560}{g} per pin/pad. This pressure, which is comparable to that used in our previous work on Pogo pins~\cite{Bejanin:2016}, will not damage the pins;
	\end{enumerate}
	\item In both bonding procedures, a~\SI{20}{\kilo\hertz} ultrasonic signal can be applied to ease pin-pad connection;
	\item Standard In~bump bonding as in flip chip is used for the ground connections at the quantum chip level. The bottom surface of the interposer block is coated with a~$\sim \SI{10}{\micro\meter}$ thick In~film that is bonded to an In~curb on the ground planes of the quantum chip [see Fig.~\ref{figure:2a:2e:mariantoni}~(d) and (e)]. The PTFE~spacers are flush with the uncoated Cu~surface of the interposer block.
\end{enumerate}

We note that, in the case of conical bonding, the pin may be coated with~In or Al, the latter being stronger and more resilient during piercing. It is well known that~Al diffuses very rapidly into~In~\cite{McRae:2018}, thus permitting an excellent electrical connection. Both Al and In films must be cleaned from their native oxide layers prior to piercing.

\section{CONCLUSION}
	\label{sec:CONCLUSION}

In the case of an interconnect with full vertical access, as for pin-chip bonding, when~$p_{\textrm{w}} / p_{\textrm{q}} \leqslant 1$ the second condition on dense wiring reduces to
\begin{equation}
N_{\textrm{q}} \equiv \left( \frac{\ell}{p_{\textrm{q}}} \right)^2 \leqslant N_{\textrm{w}} \equiv \left( \frac{\ell}{p_{\textrm{w}}} \right)^2 \, .
	\label{eq:dense:wiring:condition:2:vertical}
\end{equation}
without any constraint on~$\ell$. Considering the dimensions provided above, we find that the maximum number of qubits is limited by the qubit size; for a square chip out of a~\SI{12}{in.} wafer, $\ell = \SI{200}{\milli\meter}$ and, thus, $N_{\textrm{q}} = 160000$. This is a factor of more than~$100$ larger than for flip chip and, e.g., it will make it possible to implement~$\sim 80$ quantum error corrected qubits~\cite{Fowler:2012}, a remarkable number to start realize useful quantum algorithms.

Note that, in the case of flip chip, it is unpractical to adopt a very large chip. A~\SI{200}{\milli\meter\squared} chip, for example, would correspond to~$N_{\textrm{w}} = 4 \times \SI{200}{\milli\meter} / \SI{56}{\micro\meter} = 14286$ using optimal flip chip dimensions. In this case, the qubit pitch should be~$p_{\textrm{q}} \approx \SI{1.7}{\milli\meter}$ to fully take advantage of the chip size. Such a qubit spacing could be realized by means of qubit-resonator-qubit cells, where the resonator acts as a coherent qubit spacer. However, microwave planar resonators are bound to lengths in the few~millimeter range due to their wavelength, typically larger than~\SI{1.7}{\milli\meter}. For instance, in the IBM Q quantum chip~\cite{Bronn:2018}, the qubits are spaced by~$\approx \SI{3.5}{\milli\meter}$; in this case, we find~$N_{\textrm{q}} = ( \SI{200}{\milli\meter} )^2 / ( \SI{3.5}{\milli\meter} )^2 = 3270$. More compact coherent spacers may be realized using lumped element~$L C$ resonators or planar resonators with much higher resonance frequency. In all these cases, however, the spacer impact on qubit gate fidelities may be important.

Lastly, it is worth mentioning that rectangular coaxial ribbon cables can easily be engineered to embed attenuators and filters along the transmission lines, without impacting the wiring footprint. For example, Au or palladium traces can be added where required to form the large resistors used in attenuators. Infrared filters can be realized by altering the polyimide tape with a magnetically loaded black dye or a thin paint (or spray). Bundles of ribbon cables can be then pressed laterally in correspondence with attenuators or filters and heat sunk to the various stages of the~DR.

\begin{acknowledgments}
We thank our fruitful discussions with David P.~DiVincenzo about qubit operation at~\SI{3}{\kelvin} as well as Bluhm's group for discussions about heat shielding interposer chips.
\end{acknowledgments}


\begin{thebibliography}{35}%
\makeatletter
\providecommand \@ifxundefined [1]{%
 \@ifx{#1\undefined}
}%
\providecommand \@ifnum [1]{%
 \ifnum #1\expandafter \@firstoftwo
 \else \expandafter \@secondoftwo
 \fi
}%
\providecommand \@ifx [1]{%
 \ifx #1\expandafter \@firstoftwo
 \else \expandafter \@secondoftwo
 \fi
}%
\providecommand \natexlab [1]{#1}%
\providecommand \enquote  [1]{``#1''}%
\providecommand \bibnamefont  [1]{#1}%
\providecommand \bibfnamefont [1]{#1}%
\providecommand \citenamefont [1]{#1}%
\providecommand \href@noop [0]{\@secondoftwo}%
\providecommand \href [0]{\begingroup \@sanitize@url \@href}%
\providecommand \@href[1]{\@@startlink{#1}\@@href}%
\providecommand \@@href[1]{\endgroup#1\@@endlink}%
\providecommand \@sanitize@url [0]{\catcode `\\12\catcode `\$12\catcode
  `\&12\catcode `\#12\catcode `\^12\catcode `\_12\catcode `\%12\relax}%
\providecommand \@@startlink[1]{}%
\providecommand \@@endlink[0]{}%
\providecommand \url  [0]{\begingroup\@sanitize@url \@url }%
\providecommand \@url [1]{\endgroup\@href {#1}{\urlprefix }}%
\providecommand \urlprefix  [0]{URL }%
\providecommand \Eprint [0]{\href }%
\providecommand \doibase [0]{http://dx.doi.org/}%
\providecommand \selectlanguage [0]{\@gobble}%
\providecommand \bibinfo  [0]{\@secondoftwo}%
\providecommand \bibfield  [0]{\@secondoftwo}%
\providecommand \translation [1]{[#1]}%
\providecommand \BibitemOpen [0]{}%
\providecommand \bibitemStop [0]{}%
\providecommand \bibitemNoStop [0]{.\EOS\space}%
\providecommand \EOS [0]{\spacefactor3000\relax}%
\providecommand \BibitemShut  [1]{\csname bibitem#1\endcsname}%
\let\auto@bib@innerbib\@empty
\bibitem [{\citenamefont {Neill}\ \emph {et~al.}(2018)\citenamefont {Neill},
  \citenamefont {Roushan}, \citenamefont {Kechedzhi}, \citenamefont {Boixo},
  \citenamefont {Isakov}, \citenamefont {Smelyanskiy}, \citenamefont {Megrant},
  \citenamefont {Chiaro}, \citenamefont {Dunsworth}, \citenamefont {Arya},
  \citenamefont {Barends}, \citenamefont {Burkett}, \citenamefont {Chen},
  \citenamefont {Chen}, \citenamefont {Fowler}, \citenamefont {Foxen},
  \citenamefont {Giustina}, \citenamefont {Graff}, \citenamefont {Jeffrey},
  \citenamefont {Huang}, \citenamefont {Kelly}, \citenamefont {Klimov},
  \citenamefont {Lucero}, \citenamefont {Mutus}, \citenamefont {Neeley},
  \citenamefont {Quintana}, \citenamefont {Sank}, \citenamefont {Vainsencher},
  \citenamefont {Wenner}, \citenamefont {White}, \citenamefont {Neven},\ and\
  \citenamefont {Martinis}}]{Neill:2018}%
  \BibitemOpen
  \bibfield  {author} {\bibinfo {author} {\bibfnamefont {C.}~\bibnamefont
  {Neill}}, \bibinfo {author} {\bibfnamefont {P.}~\bibnamefont {Roushan}},
  \bibinfo {author} {\bibfnamefont {K.}~\bibnamefont {Kechedzhi}}, \bibinfo
  {author} {\bibfnamefont {S.}~\bibnamefont {Boixo}}, \bibinfo {author}
  {\bibfnamefont {S.~V.}\ \bibnamefont {Isakov}}, \bibinfo {author}
  {\bibfnamefont {V.}~\bibnamefont {Smelyanskiy}}, \bibinfo {author}
  {\bibfnamefont {A.}~\bibnamefont {Megrant}}, \bibinfo {author} {\bibfnamefont
  {B.}~\bibnamefont {Chiaro}}, \bibinfo {author} {\bibfnamefont
  {A.}~\bibnamefont {Dunsworth}}, \bibinfo {author} {\bibfnamefont
  {K.}~\bibnamefont {Arya}}, \bibinfo {author} {\bibfnamefont {R.}~\bibnamefont
  {Barends}}, \bibinfo {author} {\bibfnamefont {B.}~\bibnamefont {Burkett}},
  \bibinfo {author} {\bibfnamefont {Y.}~\bibnamefont {Chen}}, \bibinfo {author}
  {\bibfnamefont {Z.}~\bibnamefont {Chen}}, \bibinfo {author} {\bibfnamefont
  {A.}~\bibnamefont {Fowler}}, \bibinfo {author} {\bibfnamefont
  {B.}~\bibnamefont {Foxen}}, \bibinfo {author} {\bibfnamefont
  {M.}~\bibnamefont {Giustina}}, \bibinfo {author} {\bibfnamefont
  {R.}~\bibnamefont {Graff}}, \bibinfo {author} {\bibfnamefont
  {E.}~\bibnamefont {Jeffrey}}, \bibinfo {author} {\bibfnamefont
  {T.}~\bibnamefont {Huang}}, \bibinfo {author} {\bibfnamefont
  {J.}~\bibnamefont {Kelly}}, \bibinfo {author} {\bibfnamefont
  {P.}~\bibnamefont {Klimov}}, \bibinfo {author} {\bibfnamefont
  {E.}~\bibnamefont {Lucero}}, \bibinfo {author} {\bibfnamefont
  {J.}~\bibnamefont {Mutus}}, \bibinfo {author} {\bibfnamefont
  {M.}~\bibnamefont {Neeley}}, \bibinfo {author} {\bibfnamefont
  {C.}~\bibnamefont {Quintana}}, \bibinfo {author} {\bibfnamefont
  {D.}~\bibnamefont {Sank}}, \bibinfo {author} {\bibfnamefont {A.}~\bibnamefont
  {Vainsencher}}, \bibinfo {author} {\bibfnamefont {J.}~\bibnamefont {Wenner}},
  \bibinfo {author} {\bibfnamefont {T.~C.}\ \bibnamefont {White}}, \bibinfo
  {author} {\bibfnamefont {H.}~\bibnamefont {Neven}}, \ and\ \bibinfo {author}
  {\bibfnamefont {J.~M.}\ \bibnamefont {Martinis}},\ }\bibfield  {title}
  {\enquote {\bibinfo {title} {A blueprint for demonstrating quantum supremacy
  with superconducting qubits},}\ }\href {\doibase 10.1126/science.aao4309}
  {\bibfield  {journal} {\bibinfo  {journal} {Science}\ }\textbf {\bibinfo
  {volume} {360}},\ \bibinfo {pages} {195--199} (\bibinfo {year}
  {2018})}\BibitemShut {NoStop}%
\bibitem [{\citenamefont {Ladd}\ \emph {et~al.}(2010)\citenamefont {Ladd},
  \citenamefont {Jelezko}, \citenamefont {Laflamme}, \citenamefont {Nakamura},
  \citenamefont {Monroe},\ and\ \citenamefont {O'Brien}}]{Ladd:2010}%
  \BibitemOpen
  \bibfield  {author} {\bibinfo {author} {\bibfnamefont {Thaddeus~D.}\
  \bibnamefont {Ladd}}, \bibinfo {author} {\bibfnamefont {Fedor}\ \bibnamefont
  {Jelezko}}, \bibinfo {author} {\bibfnamefont {Raymond}\ \bibnamefont
  {Laflamme}}, \bibinfo {author} {\bibfnamefont {Yasunobu}\ \bibnamefont
  {Nakamura}}, \bibinfo {author} {\bibfnamefont {Christopher}\ \bibnamefont
  {Monroe}}, \ and\ \bibinfo {author} {\bibfnamefont {Jeremy~L.}\ \bibnamefont
  {O'Brien}},\ }\bibfield  {title} {\enquote {\bibinfo {title} {Quantum
  computers},}\ }\href {\doibase 10.1038/nature08812} {\bibfield  {journal}
  {\bibinfo  {journal} {Nature}\ }\textbf {\bibinfo {volume} {464}},\ \bibinfo
  {pages} {45--53} (\bibinfo {year} {2010})}\BibitemShut {NoStop}%
\bibitem [{\citenamefont {Clarke}\ and\ \citenamefont
  {Wilhelm}(2008)}]{Clarke:2008}%
  \BibitemOpen
  \bibfield  {author} {\bibinfo {author} {\bibfnamefont {John}\ \bibnamefont
  {Clarke}}\ and\ \bibinfo {author} {\bibfnamefont {Frank~K.}\ \bibnamefont
  {Wilhelm}},\ }\bibfield  {title} {\enquote {\bibinfo {title} {Superconducting
  quantum bits},}\ }\href {\doibase 10.1038/nature07128} {\bibfield  {journal}
  {\bibinfo  {journal} {Nature}\ }\textbf {\bibinfo {volume} {453}},\ \bibinfo
  {pages} {1031--1042} (\bibinfo {year} {2008})}\BibitemShut {NoStop}%
\bibitem [{\citenamefont {Wendin}(2017)}]{Wendin:2017}%
  \BibitemOpen
  \bibfield  {author} {\bibinfo {author} {\bibfnamefont {G}~\bibnamefont
  {Wendin}},\ }\bibfield  {title} {\enquote {\bibinfo {title} {Quantum
  information processing with superconducting circuits: a review},}\ }\href
  {http://stacks.iop.org/0034-4885/80/i=10/a=106001} {\bibfield  {journal}
  {\bibinfo  {journal} {Reports on Progress in Physics}\ }\textbf {\bibinfo
  {volume} {80}},\ \bibinfo {pages} {106001} (\bibinfo {year}
  {2017})}\BibitemShut {NoStop}%
\bibitem [{\citenamefont {Knight}(2017)}]{Knight:2017}%
  \BibitemOpen
  \bibfield  {author} {\bibinfo {author} {\bibfnamefont {Will}\ \bibnamefont
  {Knight}},\ }\bibfield  {title} {\enquote {\bibinfo {title} {Ibm raises the
  bar with a 50-qubit quantum computer},}\ }\href
  {https://www.technologyreview.com/s/609451/} {\bibfield  {journal} {\bibinfo
  {journal} {MIT Technology Review}\ } (\bibinfo {year} {2017})}\BibitemShut
  {NoStop}%
\bibitem [{\citenamefont {Giles}\ and\ \citenamefont
  {Knight}(2018)}]{Giles:2018}%
  \BibitemOpen
  \bibfield  {author} {\bibinfo {author} {\bibfnamefont {Martin}\ \bibnamefont
  {Giles}}\ and\ \bibinfo {author} {\bibfnamefont {Will}\ \bibnamefont
  {Knight}},\ }\bibfield  {title} {\enquote {\bibinfo {title} {Google thinks
  it’s close to “quantum supremacy.” here’s what that really means.}}\
  }\href {https://www.technologyreview.com/s/610274/} {\bibfield  {journal}
  {\bibinfo  {journal} {MIT Technology Review}\ } (\bibinfo {year}
  {2018})}\BibitemShut {NoStop}%
\bibitem [{\citenamefont {Montanaro}(2016)}]{Montanaro:2016}%
  \BibitemOpen
  \bibfield  {author} {\bibinfo {author} {\bibfnamefont {Ashley}\ \bibnamefont
  {Montanaro}},\ }\bibfield  {title} {\enquote {\bibinfo {title} {Quantum
  algorithms: An overview},}\ }\href {\doibase 10.1038/npjqi.2015.23}
  {\bibfield  {journal} {\bibinfo  {journal} {npj Quantum Information}\
  }\textbf {\bibinfo {volume} {2}},\ \bibinfo {pages} {15023} (\bibinfo {year}
  {2016})}\BibitemShut {NoStop}%
\bibitem [{\citenamefont {Gottesman}(2010)}]{Gottesman:2010}%
  \BibitemOpen
  \bibfield  {author} {\bibinfo {author} {\bibfnamefont {Daniel}\ \bibnamefont
  {Gottesman}},\ }\bibfield  {title} {\enquote {\bibinfo {title} {An
  introduction to quantum error correction and fault-tolerant quantum
  computation},}\ }\href {\doibase 10.1090/psapm/068/2762145} {\bibfield
  {journal} {\bibinfo  {journal} {Quantum Information Science and Its
  Contributions to Mathematics, Proceedings of Symposia in Applied
  Mathematics}\ }\textbf {\bibinfo {volume} {68}},\ \bibinfo {pages} {13--58}
  (\bibinfo {year} {2010})}\BibitemShut {NoStop}%
\bibitem [{\citenamefont {Nielsen}\ and\ \citenamefont
  {Chuang}(2000)}]{Nielsen:2000}%
  \BibitemOpen
  \bibfield  {author} {\bibinfo {author} {\bibfnamefont {Michael~A.}\
  \bibnamefont {Nielsen}}\ and\ \bibinfo {author} {\bibfnamefont {Isaac~L.}\
  \bibnamefont {Chuang}},\ }\href {\doibase 10.1017/CBO9780511976667} {\emph
  {\bibinfo {title} {Quantum Computation and Quantum Information}}}\ (\bibinfo
  {publisher} {Cambridge University Press ({CUP})},\ \bibinfo {address}
  {Cambridge, {UK}},\ \bibinfo {year} {2000})\BibitemShut {NoStop}%
\bibitem [{\citenamefont {Koch}\ \emph {et~al.}(2007)\citenamefont {Koch},
  \citenamefont {DiVincenzo},\ and\ \citenamefont {Clarke}}]{Koch:2007}%
  \BibitemOpen
  \bibfield  {author} {\bibinfo {author} {\bibfnamefont {Roger~H.}\
  \bibnamefont {Koch}}, \bibinfo {author} {\bibfnamefont {David~P.}\
  \bibnamefont {DiVincenzo}}, \ and\ \bibinfo {author} {\bibfnamefont {John}\
  \bibnamefont {Clarke}},\ }\bibfield  {title} {\enquote {\bibinfo {title}
  {Model for $1/f$ flux noise in squids and qubits},}\ }\href {\doibase
  10.1103/PhysRevLett.98.267003} {\bibfield  {journal} {\bibinfo  {journal}
  {Phys. Rev. Lett.}\ }\textbf {\bibinfo {volume} {98}},\ \bibinfo {pages}
  {267003} (\bibinfo {year} {2007})}\BibitemShut {NoStop}%
\bibitem [{\citenamefont {Barends}\ \emph {et~al.}(2013)\citenamefont
  {Barends}, \citenamefont {Kelly}, \citenamefont {Megrant}, \citenamefont
  {Sank}, \citenamefont {Jeffrey}, \citenamefont {Chen}, \citenamefont {Yin},
  \citenamefont {Chiaro}, \citenamefont {Mutus}, \citenamefont {Neill},
  \citenamefont {O'Malley}, \citenamefont {Roushan}, \citenamefont {Wenner},
  \citenamefont {White}, \citenamefont {Cleland},\ and\ \citenamefont
  {Martinis}}]{Barends:2013}%
  \BibitemOpen
  \bibfield  {author} {\bibinfo {author} {\bibfnamefont {Rami}\ \bibnamefont
  {Barends}}, \bibinfo {author} {\bibfnamefont {Julian}\ \bibnamefont {Kelly}},
  \bibinfo {author} {\bibfnamefont {Anthony}\ \bibnamefont {Megrant}}, \bibinfo
  {author} {\bibfnamefont {Daniel}\ \bibnamefont {Sank}}, \bibinfo {author}
  {\bibfnamefont {Evan}\ \bibnamefont {Jeffrey}}, \bibinfo {author}
  {\bibfnamefont {Yu}~\bibnamefont {Chen}}, \bibinfo {author} {\bibfnamefont
  {Yi}~\bibnamefont {Yin}}, \bibinfo {author} {\bibfnamefont {Ben}\
  \bibnamefont {Chiaro}}, \bibinfo {author} {\bibfnamefont {Josh}\ \bibnamefont
  {Mutus}}, \bibinfo {author} {\bibfnamefont {Charles}\ \bibnamefont {Neill}},
  \bibinfo {author} {\bibfnamefont {Peter}\ \bibnamefont {O'Malley}}, \bibinfo
  {author} {\bibfnamefont {Pedram}\ \bibnamefont {Roushan}}, \bibinfo {author}
  {\bibfnamefont {James}\ \bibnamefont {Wenner}}, \bibinfo {author}
  {\bibfnamefont {Theodor~C.}\ \bibnamefont {White}}, \bibinfo {author}
  {\bibfnamefont {Aandrew~N.}\ \bibnamefont {Cleland}}, \ and\ \bibinfo
  {author} {\bibfnamefont {John~M.}\ \bibnamefont {Martinis}},\ }\bibfield
  {title} {\enquote {\bibinfo {title} {Coherent josephson qubit suitable for
  scalable quantum integrated circuits},}\ }\href {\doibase
  10.1103/PhysRevLett.111.080502} {\bibfield  {journal} {\bibinfo  {journal}
  {Phys. Rev. Lett.}\ }\textbf {\bibinfo {volume} {111}},\ \bibinfo {pages}
  {080502} (\bibinfo {year} {2013})}\BibitemShut {NoStop}%
\bibitem [{\citenamefont {Kelly}\ \emph {et~al.}(2015)\citenamefont {Kelly},
  \citenamefont {Barends}, \citenamefont {Fowler}, \citenamefont {Megrant},
  \citenamefont {Jeffrey}, \citenamefont {White}, \citenamefont {Sank},
  \citenamefont {Mutus}, \citenamefont {Campbell}, \citenamefont {Chen},
  \citenamefont {Chen}, \citenamefont {Chiaro}, \citenamefont {Dunsworth},
  \citenamefont {Hoi}, \citenamefont {Neill}, \citenamefont {O'Malley},
  \citenamefont {Quintana}, \citenamefont {Roushan}, \citenamefont
  {Vainsencher}, \citenamefont {Wenner}, \citenamefont {Cleland},\ and\
  \citenamefont {Martinis}}]{Kelly:2015}%
  \BibitemOpen
  \bibfield  {author} {\bibinfo {author} {\bibfnamefont {Julian}\ \bibnamefont
  {Kelly}}, \bibinfo {author} {\bibfnamefont {Rami}\ \bibnamefont {Barends}},
  \bibinfo {author} {\bibfnamefont {Austin~G.}\ \bibnamefont {Fowler}},
  \bibinfo {author} {\bibfnamefont {Anthony}\ \bibnamefont {Megrant}}, \bibinfo
  {author} {\bibfnamefont {Evan}\ \bibnamefont {Jeffrey}}, \bibinfo {author}
  {\bibfnamefont {Theodore~C.}\ \bibnamefont {White}}, \bibinfo {author}
  {\bibfnamefont {Daniel}\ \bibnamefont {Sank}}, \bibinfo {author}
  {\bibfnamefont {Josh~Y.}\ \bibnamefont {Mutus}}, \bibinfo {author}
  {\bibfnamefont {Brooks}\ \bibnamefont {Campbell}}, \bibinfo {author}
  {\bibfnamefont {Yu}~\bibnamefont {Chen}}, \bibinfo {author} {\bibfnamefont
  {Zijun}\ \bibnamefont {Chen}}, \bibinfo {author} {\bibfnamefont {Ben}\
  \bibnamefont {Chiaro}}, \bibinfo {author} {\bibfnamefont {Andrew}\
  \bibnamefont {Dunsworth}}, \bibinfo {author} {\bibfnamefont {Io-Chun}\
  \bibnamefont {Hoi}}, \bibinfo {author} {\bibfnamefont {Charles}\ \bibnamefont
  {Neill}}, \bibinfo {author} {\bibfnamefont {Peter~J.J.}\ \bibnamefont
  {O'Malley}}, \bibinfo {author} {\bibfnamefont {Christopher}\ \bibnamefont
  {Quintana}}, \bibinfo {author} {\bibfnamefont {Pedram}\ \bibnamefont
  {Roushan}}, \bibinfo {author} {\bibfnamefont {Amit}\ \bibnamefont
  {Vainsencher}}, \bibinfo {author} {\bibfnamefont {James}\ \bibnamefont
  {Wenner}}, \bibinfo {author} {\bibfnamefont {Andrew~N.}\ \bibnamefont
  {Cleland}}, \ and\ \bibinfo {author} {\bibfnamefont {John~M.}\ \bibnamefont
  {Martinis}},\ }\bibfield  {title} {\enquote {\bibinfo {title} {State
  preservation by repetitive error detection in a superconducting quantum
  circuit},}\ }\href {\doibase 10.1038/nature14270} {\bibfield  {journal}
  {\bibinfo  {journal} {Nature}\ }\textbf {\bibinfo {volume} {519}},\ \bibinfo
  {pages} {66--69} (\bibinfo {year} {2015})}\BibitemShut {NoStop}%
\bibitem [{\citenamefont {Versluis}\ \emph {et~al.}(2017)\citenamefont
  {Versluis}, \citenamefont {Poletto}, \citenamefont {Khammassi}, \citenamefont
  {Tarasinski}, \citenamefont {Haider}, \citenamefont {Michalak}, \citenamefont
  {Bruno}, \citenamefont {Bertels},\ and\ \citenamefont
  {DiCarlo}}]{Versluis:2017}%
  \BibitemOpen
  \bibfield  {author} {\bibinfo {author} {\bibfnamefont {R.}~\bibnamefont
  {Versluis}}, \bibinfo {author} {\bibfnamefont {S.}~\bibnamefont {Poletto}},
  \bibinfo {author} {\bibfnamefont {N.}~\bibnamefont {Khammassi}}, \bibinfo
  {author} {\bibfnamefont {B.}~\bibnamefont {Tarasinski}}, \bibinfo {author}
  {\bibfnamefont {N.}~\bibnamefont {Haider}}, \bibinfo {author} {\bibfnamefont
  {D.~J.}\ \bibnamefont {Michalak}}, \bibinfo {author} {\bibfnamefont
  {A.}~\bibnamefont {Bruno}}, \bibinfo {author} {\bibfnamefont
  {K.}~\bibnamefont {Bertels}}, \ and\ \bibinfo {author} {\bibfnamefont
  {L.}~\bibnamefont {DiCarlo}},\ }\bibfield  {title} {\enquote {\bibinfo
  {title} {Scalable quantum circuit and control for a superconducting surface
  code},}\ }\href {\doibase 10.1103/PhysRevApplied.8.034021} {\bibfield
  {journal} {\bibinfo  {journal} {Phys. Rev. Applied}\ }\textbf {\bibinfo
  {volume} {8}},\ \bibinfo {pages} {034021} (\bibinfo {year}
  {2017})}\BibitemShut {NoStop}%
\bibitem [{\citenamefont {Otterbach}\ \emph {et~al.}(2017)\citenamefont
  {Otterbach}, \citenamefont {Manenti}, \citenamefont {Alidoust}, \citenamefont
  {Bestwick}, \citenamefont {Block}, \citenamefont {Bloom}, \citenamefont
  {Caldwell}, \citenamefont {Didier}, \citenamefont {Fried}, \citenamefont
  {Hong}, \citenamefont {Karalekas}, \citenamefont {Osborn}, \citenamefont
  {Papageorge}, \citenamefont {Peterson}, \citenamefont {Prawiroatmodjo},
  \citenamefont {Rubin}, \citenamefont {Ryan}, \citenamefont {Scarabelli},
  \citenamefont {Scheer}, \citenamefont {Sete}, \citenamefont {Sivarajah},
  \citenamefont {Smith}, \citenamefont {Staley}, \citenamefont {Tezak},
  \citenamefont {Zeng}, \citenamefont {Hudson}, \citenamefont {Johnson},
  \citenamefont {Reagor}, \citenamefont {da~Silva},\ and\ \citenamefont
  {Rigetti}}]{Otterbach:2017}%
  \BibitemOpen
  \bibfield  {author} {\bibinfo {author} {\bibfnamefont {J.~S.}\ \bibnamefont
  {Otterbach}}, \bibinfo {author} {\bibfnamefont {R.}~\bibnamefont {Manenti}},
  \bibinfo {author} {\bibfnamefont {N.}~\bibnamefont {Alidoust}}, \bibinfo
  {author} {\bibfnamefont {A.}~\bibnamefont {Bestwick}}, \bibinfo {author}
  {\bibfnamefont {M.}~\bibnamefont {Block}}, \bibinfo {author} {\bibfnamefont
  {B.}~\bibnamefont {Bloom}}, \bibinfo {author} {\bibfnamefont
  {S.}~\bibnamefont {Caldwell}}, \bibinfo {author} {\bibfnamefont
  {N.}~\bibnamefont {Didier}}, \bibinfo {author} {\bibfnamefont {E.~Schuyler}\
  \bibnamefont {Fried}}, \bibinfo {author} {\bibfnamefont {S.}~\bibnamefont
  {Hong}}, \bibinfo {author} {\bibfnamefont {P.}~\bibnamefont {Karalekas}},
  \bibinfo {author} {\bibfnamefont {C.~B.}\ \bibnamefont {Osborn}}, \bibinfo
  {author} {\bibfnamefont {A.}~\bibnamefont {Papageorge}}, \bibinfo {author}
  {\bibfnamefont {E.~C.}\ \bibnamefont {Peterson}}, \bibinfo {author}
  {\bibfnamefont {G.}~\bibnamefont {Prawiroatmodjo}}, \bibinfo {author}
  {\bibfnamefont {N.}~\bibnamefont {Rubin}}, \bibinfo {author} {\bibfnamefont
  {Colm~A.}\ \bibnamefont {Ryan}}, \bibinfo {author} {\bibfnamefont
  {D.}~\bibnamefont {Scarabelli}}, \bibinfo {author} {\bibfnamefont
  {M.}~\bibnamefont {Scheer}}, \bibinfo {author} {\bibfnamefont {E.~A.}\
  \bibnamefont {Sete}}, \bibinfo {author} {\bibfnamefont {P.}~\bibnamefont
  {Sivarajah}}, \bibinfo {author} {\bibfnamefont {Robert~S.}\ \bibnamefont
  {Smith}}, \bibinfo {author} {\bibfnamefont {A.}~\bibnamefont {Staley}},
  \bibinfo {author} {\bibfnamefont {N.}~\bibnamefont {Tezak}}, \bibinfo
  {author} {\bibfnamefont {W.~J.}\ \bibnamefont {Zeng}}, \bibinfo {author}
  {\bibfnamefont {A.}~\bibnamefont {Hudson}}, \bibinfo {author} {\bibfnamefont
  {Blake~R.}\ \bibnamefont {Johnson}}, \bibinfo {author} {\bibfnamefont
  {M.}~\bibnamefont {Reagor}}, \bibinfo {author} {\bibfnamefont {M.~P.}\
  \bibnamefont {da~Silva}}, \ and\ \bibinfo {author} {\bibfnamefont
  {C.}~\bibnamefont {Rigetti}},\ }\href {https://arxiv.org/abs/1712.05771}
  {\enquote {\bibinfo {title} {Unsupervised machine learning on a hybrid
  quantum computer},}\ } (\bibinfo {year} {2017}),\ \Eprint
  {http://arxiv.org/abs/arXiv:1712.05771} {arXiv:1712.05771} \BibitemShut
  {NoStop}%
\bibitem [{\citenamefont {Martinis}(2015)}]{Martinis:2015}%
  \BibitemOpen
  \bibfield  {author} {\bibinfo {author} {\bibfnamefont {John~M.}\ \bibnamefont
  {Martinis}},\ }\bibfield  {title} {\enquote {\bibinfo {title} {Qubit
  metrology for building a fault-tolerant quantum computer},}\ }\href {\doibase
  10.1038/npjqi.2015.5} {\bibfield  {journal} {\bibinfo  {journal} {npj Quantum
  Information}\ }\textbf {\bibinfo {volume} {1}},\ \bibinfo {pages} {15005}
  (\bibinfo {year} {2015})}\BibitemShut {NoStop}%
\bibitem [{\citenamefont {Tuckerman}\ \emph {et~al.}(2016)\citenamefont
  {Tuckerman}, \citenamefont {Hamilton}, \citenamefont {Reilly}, \citenamefont
  {Bai}, \citenamefont {Hernandez}, \citenamefont {Hornibrook}, \citenamefont
  {Sellers},\ and\ \citenamefont {Ellis}}]{Tuckerman:2016}%
  \BibitemOpen
  \bibfield  {author} {\bibinfo {author} {\bibfnamefont {David~B}\ \bibnamefont
  {Tuckerman}}, \bibinfo {author} {\bibfnamefont {Michael~C}\ \bibnamefont
  {Hamilton}}, \bibinfo {author} {\bibfnamefont {David~J}\ \bibnamefont
  {Reilly}}, \bibinfo {author} {\bibfnamefont {Rujun}\ \bibnamefont {Bai}},
  \bibinfo {author} {\bibfnamefont {George~A}\ \bibnamefont {Hernandez}},
  \bibinfo {author} {\bibfnamefont {John~M}\ \bibnamefont {Hornibrook}},
  \bibinfo {author} {\bibfnamefont {John~A}\ \bibnamefont {Sellers}}, \ and\
  \bibinfo {author} {\bibfnamefont {Charles~D}\ \bibnamefont {Ellis}},\
  }\bibfield  {title} {\enquote {\bibinfo {title} {Flexible superconducting nb
  transmission lines on thin film polyimide for quantum computing
  applications},}\ }\href {http://stacks.iop.org/0953-2048/29/i=8/a=084007}
  {\bibfield  {journal} {\bibinfo  {journal} {Superconductor Science and
  Technology}\ }\textbf {\bibinfo {volume} {29}},\ \bibinfo {pages} {084007}
  (\bibinfo {year} {2016})}\BibitemShut {NoStop}%
\bibitem [{\citenamefont {B\'ejanin}\ \emph {et~al.}(2016)\citenamefont
  {B\'ejanin}, \citenamefont {McConkey}, \citenamefont {Rinehart},
  \citenamefont {Earnest}, \citenamefont {McRae}, \citenamefont {Shiri},
  \citenamefont {Bateman}, \citenamefont {Rohanizadegan}, \citenamefont
  {Penava}, \citenamefont {Breul}, \citenamefont {Royak}, \citenamefont
  {Zapatka}, \citenamefont {Fowler},\ and\ \citenamefont
  {Mariantoni}}]{Bejanin:2016}%
  \BibitemOpen
  \bibfield  {author} {\bibinfo {author} {\bibfnamefont {J.~H.}\ \bibnamefont
  {B\'ejanin}}, \bibinfo {author} {\bibfnamefont {T.~G.}\ \bibnamefont
  {McConkey}}, \bibinfo {author} {\bibfnamefont {J.~R.}\ \bibnamefont
  {Rinehart}}, \bibinfo {author} {\bibfnamefont {C.~T.}\ \bibnamefont
  {Earnest}}, \bibinfo {author} {\bibfnamefont {C.~R.~H.}\ \bibnamefont
  {McRae}}, \bibinfo {author} {\bibfnamefont {D.}~\bibnamefont {Shiri}},
  \bibinfo {author} {\bibfnamefont {J.~D.}\ \bibnamefont {Bateman}}, \bibinfo
  {author} {\bibfnamefont {Y.}~\bibnamefont {Rohanizadegan}}, \bibinfo {author}
  {\bibfnamefont {B.}~\bibnamefont {Penava}}, \bibinfo {author} {\bibfnamefont
  {P.}~\bibnamefont {Breul}}, \bibinfo {author} {\bibfnamefont
  {S.}~\bibnamefont {Royak}}, \bibinfo {author} {\bibfnamefont
  {M.}~\bibnamefont {Zapatka}}, \bibinfo {author} {\bibfnamefont {A.~G.}\
  \bibnamefont {Fowler}}, \ and\ \bibinfo {author} {\bibfnamefont
  {M.}~\bibnamefont {Mariantoni}},\ }\bibfield  {title} {\enquote {\bibinfo
  {title} {Three-dimensional wiring for extensible quantum computing: The
  quantum socket},}\ }\href {\doibase 10.1103/PhysRevApplied.6.044010}
  {\bibfield  {journal} {\bibinfo  {journal} {Phys. Rev. Applied}\ }\textbf
  {\bibinfo {volume} {6}},\ \bibinfo {pages} {044010} (\bibinfo {year}
  {2016})}\BibitemShut {NoStop}%
\bibitem [{\citenamefont {Wenner}\ \emph {et~al.}(2011)\citenamefont {Wenner},
  \citenamefont {Neeley}, \citenamefont {Bialczak}, \citenamefont {Lenander},
  \citenamefont {Lucero}, \citenamefont {O'Connell}, \citenamefont {Sank},
  \citenamefont {Wang}, \citenamefont {Weides}, \citenamefont {Cleland},\ and\
  \citenamefont {Martinis}}]{Wenner:2011:a}%
  \BibitemOpen
  \bibfield  {author} {\bibinfo {author} {\bibfnamefont {James}\ \bibnamefont
  {Wenner}}, \bibinfo {author} {\bibfnamefont {Matthew}\ \bibnamefont
  {Neeley}}, \bibinfo {author} {\bibfnamefont {Radoslaw~C.}\ \bibnamefont
  {Bialczak}}, \bibinfo {author} {\bibfnamefont {Michael}\ \bibnamefont
  {Lenander}}, \bibinfo {author} {\bibfnamefont {Erik}\ \bibnamefont {Lucero}},
  \bibinfo {author} {\bibfnamefont {Aaron~D.}\ \bibnamefont {O'Connell}},
  \bibinfo {author} {\bibfnamefont {Daniel}\ \bibnamefont {Sank}}, \bibinfo
  {author} {\bibfnamefont {Haohua}\ \bibnamefont {Wang}}, \bibinfo {author}
  {\bibfnamefont {Martin}\ \bibnamefont {Weides}}, \bibinfo {author}
  {\bibfnamefont {Andrew~N.}\ \bibnamefont {Cleland}}, \ and\ \bibinfo {author}
  {\bibfnamefont {John~M.}\ \bibnamefont {Martinis}},\ }\bibfield  {title}
  {\enquote {\bibinfo {title} {Wirebond crosstalk and cavity modes in large
  chip mounts for superconducting qubits},}\ }\href {\doibase
  10.1088/0953-2048/24/6/065001} {\bibfield  {journal} {\bibinfo  {journal}
  {Superconductor Science and Technology}\ }\textbf {\bibinfo {volume} {24}},\
  \bibinfo {pages} {065001} (\bibinfo {year} {2011})}\BibitemShut {NoStop}%
\bibitem [{\citenamefont {Earnest}\ \emph {et~al.}(2018)\citenamefont
  {Earnest}, \citenamefont {B\'ejanin}, \citenamefont {McConkey}, \citenamefont
  {Peters}, \citenamefont {Korinek}, \citenamefont {Yuan},\ and\ \citenamefont
  {Mariantoni}}]{Earnest:2018}%
  \BibitemOpen
  \bibfield  {author} {\bibinfo {author} {\bibfnamefont {Carolyn~T}\
  \bibnamefont {Earnest}}, \bibinfo {author} {\bibfnamefont {J\'er\'emy~H}\
  \bibnamefont {B\'ejanin}}, \bibinfo {author} {\bibfnamefont {Thomas~G}\
  \bibnamefont {McConkey}}, \bibinfo {author} {\bibfnamefont {Evan~A.}\
  \bibnamefont {Peters}}, \bibinfo {author} {\bibfnamefont {Andreas}\
  \bibnamefont {Korinek}}, \bibinfo {author} {\bibfnamefont {Hui}\ \bibnamefont
  {Yuan}}, \ and\ \bibinfo {author} {\bibfnamefont {Matteo}\ \bibnamefont
  {Mariantoni}},\ }\bibfield  {title} {\enquote {\bibinfo {title} {Substrate
  surface engineering for high-quality silicon/aluminum superconducting
  resonators},}\ }\href {http://iopscience.iop.org/10.1088/1361-6668/aae548}
  {\bibfield  {journal} {\bibinfo  {journal} {Superconductor Science and
  Technology}\ } (\bibinfo {year} {2018})}\BibitemShut {NoStop}%
\bibitem [{\citenamefont {Bronn}\ \emph {et~al.}(2018)\citenamefont {Bronn},
  \citenamefont {Adiga}, \citenamefont {Olivadese}, \citenamefont {Wu},
  \citenamefont {Chow},\ and\ \citenamefont {Pappas}}]{Bronn:2018}%
  \BibitemOpen
  \bibfield  {author} {\bibinfo {author} {\bibfnamefont {Nicholas~T}\
  \bibnamefont {Bronn}}, \bibinfo {author} {\bibfnamefont {Vivekananda~P}\
  \bibnamefont {Adiga}}, \bibinfo {author} {\bibfnamefont {Salvatore~B}\
  \bibnamefont {Olivadese}}, \bibinfo {author} {\bibfnamefont {Xian}\
  \bibnamefont {Wu}}, \bibinfo {author} {\bibfnamefont {Jerry~M}\ \bibnamefont
  {Chow}}, \ and\ \bibinfo {author} {\bibfnamefont {David~P}\ \bibnamefont
  {Pappas}},\ }\bibfield  {title} {\enquote {\bibinfo {title} {High coherence
  plane breaking packaging for superconducting qubits},}\ }\href
  {http://stacks.iop.org/2058-9565/3/i=2/a=024007} {\bibfield  {journal}
  {\bibinfo  {journal} {Quantum Science and Technology}\ }\textbf {\bibinfo
  {volume} {3}},\ \bibinfo {pages} {024007} (\bibinfo {year}
  {2018})}\BibitemShut {NoStop}%
\bibitem [{\citenamefont {Rosenberg}\ \emph {et~al.}(2017)\citenamefont
  {Rosenberg}, \citenamefont {Kim}, \citenamefont {Das}, \citenamefont {Yost},
  \citenamefont {Gustavsson}, \citenamefont {Hover}, \citenamefont {Krantz},
  \citenamefont {Melville}, \citenamefont {Racz}, \citenamefont {Samach},
  \citenamefont {Weber}, \citenamefont {Yan}, \citenamefont {Yoder},
  \citenamefont {Kerman},\ and\ \citenamefont {Oliver}}]{Rosenberg:2017}%
  \BibitemOpen
  \bibfield  {author} {\bibinfo {author} {\bibfnamefont {Danna}\ \bibnamefont
  {Rosenberg}}, \bibinfo {author} {\bibfnamefont {David}\ \bibnamefont {Kim}},
  \bibinfo {author} {\bibfnamefont {Rabindra}\ \bibnamefont {Das}}, \bibinfo
  {author} {\bibfnamefont {Donna-Ruth}\ \bibnamefont {Yost}}, \bibinfo {author}
  {\bibfnamefont {Simon}\ \bibnamefont {Gustavsson}}, \bibinfo {author}
  {\bibfnamefont {David}\ \bibnamefont {Hover}}, \bibinfo {author}
  {\bibfnamefont {Philip}\ \bibnamefont {Krantz}}, \bibinfo {author}
  {\bibfnamefont {A.}~\bibnamefont {Melville}}, \bibinfo {author}
  {\bibfnamefont {Livia}\ \bibnamefont {Racz}}, \bibinfo {author}
  {\bibfnamefont {Gabriel~O.}\ \bibnamefont {Samach}}, \bibinfo {author}
  {\bibfnamefont {Steven~J.}\ \bibnamefont {Weber}}, \bibinfo {author}
  {\bibfnamefont {Fei}\ \bibnamefont {Yan}}, \bibinfo {author} {\bibfnamefont
  {Jonilyn}\ \bibnamefont {Yoder}}, \bibinfo {author} {\bibfnamefont
  {Andrew~J.}\ \bibnamefont {Kerman}}, \ and\ \bibinfo {author} {\bibfnamefont
  {William~D.}\ \bibnamefont {Oliver}},\ }\bibfield  {title} {\enquote
  {\bibinfo {title} {Building logical qubits in a superconducting quantum
  computing system},}\ }\href {\doibase doi:10.1038/s41534-017-0044-0}
  {\bibfield  {journal} {\bibinfo  {journal} {npj Quantum Information}\
  }\textbf {\bibinfo {volume} {3}} (\bibinfo {year} {2017}),\
  doi:10.1038/s41534-017-0044-0}\BibitemShut {NoStop}%
\bibitem [{\citenamefont {Foxen}\ \emph {et~al.}(2018)\citenamefont {Foxen},
  \citenamefont {Mutus}, \citenamefont {Lucero}, \citenamefont {Graff},
  \citenamefont {Megrant}, \citenamefont {Chen}, \citenamefont {Quintana},
  \citenamefont {Burkett}, \citenamefont {Kelly}, \citenamefont {Jeffrey},
  \citenamefont {Yang}, \citenamefont {Yu}, \citenamefont {Arya}, \citenamefont
  {Barends}, \citenamefont {Chen}, \citenamefont {Chiaro}, \citenamefont
  {Dunsworth}, \citenamefont {Fowler}, \citenamefont {Gidney}, \citenamefont
  {Giustina}, \citenamefont {Huang}, \citenamefont {Klimov}, \citenamefont
  {Neeley}, \citenamefont {Neill}, \citenamefont {Roushan}, \citenamefont
  {Sank}, \citenamefont {Vainsencher}, \citenamefont {Wenner}, \citenamefont
  {White},\ and\ \citenamefont {Martinis}}]{Foxen:2017}%
  \BibitemOpen
  \bibfield  {author} {\bibinfo {author} {\bibfnamefont {B}~\bibnamefont
  {Foxen}}, \bibinfo {author} {\bibfnamefont {J~Y}\ \bibnamefont {Mutus}},
  \bibinfo {author} {\bibfnamefont {E}~\bibnamefont {Lucero}}, \bibinfo
  {author} {\bibfnamefont {R}~\bibnamefont {Graff}}, \bibinfo {author}
  {\bibfnamefont {A}~\bibnamefont {Megrant}}, \bibinfo {author} {\bibfnamefont
  {Yu}~\bibnamefont {Chen}}, \bibinfo {author} {\bibfnamefont {C}~\bibnamefont
  {Quintana}}, \bibinfo {author} {\bibfnamefont {B}~\bibnamefont {Burkett}},
  \bibinfo {author} {\bibfnamefont {J}~\bibnamefont {Kelly}}, \bibinfo {author}
  {\bibfnamefont {E}~\bibnamefont {Jeffrey}}, \bibinfo {author} {\bibfnamefont
  {Yan}\ \bibnamefont {Yang}}, \bibinfo {author} {\bibfnamefont {Anthony}\
  \bibnamefont {Yu}}, \bibinfo {author} {\bibfnamefont {K}~\bibnamefont
  {Arya}}, \bibinfo {author} {\bibfnamefont {R}~\bibnamefont {Barends}},
  \bibinfo {author} {\bibfnamefont {Zijun}\ \bibnamefont {Chen}}, \bibinfo
  {author} {\bibfnamefont {B}~\bibnamefont {Chiaro}}, \bibinfo {author}
  {\bibfnamefont {A}~\bibnamefont {Dunsworth}}, \bibinfo {author}
  {\bibfnamefont {A}~\bibnamefont {Fowler}}, \bibinfo {author} {\bibfnamefont
  {C}~\bibnamefont {Gidney}}, \bibinfo {author} {\bibfnamefont {M}~\bibnamefont
  {Giustina}}, \bibinfo {author} {\bibfnamefont {T}~\bibnamefont {Huang}},
  \bibinfo {author} {\bibfnamefont {P}~\bibnamefont {Klimov}}, \bibinfo
  {author} {\bibfnamefont {M}~\bibnamefont {Neeley}}, \bibinfo {author}
  {\bibfnamefont {C}~\bibnamefont {Neill}}, \bibinfo {author} {\bibfnamefont
  {P}~\bibnamefont {Roushan}}, \bibinfo {author} {\bibfnamefont
  {D}~\bibnamefont {Sank}}, \bibinfo {author} {\bibfnamefont {A}~\bibnamefont
  {Vainsencher}}, \bibinfo {author} {\bibfnamefont {J}~\bibnamefont {Wenner}},
  \bibinfo {author} {\bibfnamefont {T~C}\ \bibnamefont {White}}, \ and\
  \bibinfo {author} {\bibfnamefont {John~M}\ \bibnamefont {Martinis}},\
  }\bibfield  {title} {\enquote {\bibinfo {title} {Qubit compatible
  superconducting interconnects},}\ }\href
  {http://stacks.iop.org/2058-9565/3/i=1/a=014005} {\bibfield  {journal}
  {\bibinfo  {journal} {Quantum Science and Technology}\ }\textbf {\bibinfo
  {volume} {3}},\ \bibinfo {pages} {014005} (\bibinfo {year}
  {2018})}\BibitemShut {NoStop}%
\bibitem [{\citenamefont {Das}\ \emph {et~al.}(2018)\citenamefont {Das},
  \citenamefont {Yoder}, \citenamefont {Rosenberg}, \citenamefont {Kim},
  \citenamefont {Yost}, \citenamefont {Mallek}, \citenamefont {Hover},
  \citenamefont {Bolkhovsky}, \citenamefont {Kerman},\ and\ \citenamefont
  {Oliver}}]{Das:2018}%
  \BibitemOpen
  \bibfield  {author} {\bibinfo {author} {\bibfnamefont {R.}~\bibnamefont
  {Das}}, \bibinfo {author} {\bibfnamefont {J.}~\bibnamefont {Yoder}}, \bibinfo
  {author} {\bibfnamefont {D.}~\bibnamefont {Rosenberg}}, \bibinfo {author}
  {\bibfnamefont {D.}~\bibnamefont {Kim}}, \bibinfo {author} {\bibfnamefont
  {D.}~\bibnamefont {Yost}}, \bibinfo {author} {\bibfnamefont {J.}~\bibnamefont
  {Mallek}}, \bibinfo {author} {\bibfnamefont {D.}~\bibnamefont {Hover}},
  \bibinfo {author} {\bibfnamefont {V.}~\bibnamefont {Bolkhovsky}}, \bibinfo
  {author} {\bibfnamefont {A.}~\bibnamefont {Kerman}}, \ and\ \bibinfo {author}
  {\bibfnamefont {W.}~\bibnamefont {Oliver}},\ }\bibfield  {title} {\enquote
  {\bibinfo {title} {Cryogenic qubit integration for quantum computing},}\ }in\
  \href {\doibase 10.1109/ECTC.2018.00080} {\emph {\bibinfo {booktitle} {2018
  IEEE 68th Electronic Components and Technology Conference (ECTC)}}}\
  (\bibinfo {year} {2018})\ pp.\ \bibinfo {pages} {504--514}\BibitemShut
  {NoStop}%
\bibitem [{\citenamefont {McDermott}\ \emph {et~al.}(2018)\citenamefont
  {McDermott}, \citenamefont {Vavilov}, \citenamefont {Plourde}, \citenamefont
  {Wilhelm}, \citenamefont {Liebermann}, \citenamefont {Mukhanov},\ and\
  \citenamefont {Ohki}}]{McDermott:2018}%
  \BibitemOpen
  \bibfield  {author} {\bibinfo {author} {\bibfnamefont {R}~\bibnamefont
  {McDermott}}, \bibinfo {author} {\bibfnamefont {M~G}\ \bibnamefont
  {Vavilov}}, \bibinfo {author} {\bibfnamefont {B~L~T}\ \bibnamefont
  {Plourde}}, \bibinfo {author} {\bibfnamefont {F~K}\ \bibnamefont {Wilhelm}},
  \bibinfo {author} {\bibfnamefont {P~J}\ \bibnamefont {Liebermann}}, \bibinfo
  {author} {\bibfnamefont {O~A}\ \bibnamefont {Mukhanov}}, \ and\ \bibinfo
  {author} {\bibfnamefont {T~A}\ \bibnamefont {Ohki}},\ }\bibfield  {title}
  {\enquote {\bibinfo {title} {Quantum–classical interface based on single
  flux quantum digital logic},}\ }\href
  {http://stacks.iop.org/2058-9565/3/i=2/a=024004} {\bibfield  {journal}
  {\bibinfo  {journal} {Quantum Science and Technology}\ }\textbf {\bibinfo
  {volume} {3}},\ \bibinfo {pages} {024004} (\bibinfo {year}
  {2018})}\BibitemShut {NoStop}%
\bibitem [{\citenamefont {Patra}\ \emph {et~al.}(2018)\citenamefont {Patra},
  \citenamefont {Incandela}, \citenamefont {van Dijk}, \citenamefont {Homulle},
  \citenamefont {Song}, \citenamefont {Shahmohammadi}, \citenamefont
  {Staszewski}, \citenamefont {Vladimirescu}, \citenamefont {Babaie},
  \citenamefont {Sebastiano},\ and\ \citenamefont {Charbon}}]{Patra:2018}%
  \BibitemOpen
  \bibfield  {author} {\bibinfo {author} {\bibfnamefont {B.}~\bibnamefont
  {Patra}}, \bibinfo {author} {\bibfnamefont {R.~M.}\ \bibnamefont
  {Incandela}}, \bibinfo {author} {\bibfnamefont {J.~P.~G.}\ \bibnamefont {van
  Dijk}}, \bibinfo {author} {\bibfnamefont {H.~A.~R.}\ \bibnamefont {Homulle}},
  \bibinfo {author} {\bibfnamefont {L.}~\bibnamefont {Song}}, \bibinfo {author}
  {\bibfnamefont {M.}~\bibnamefont {Shahmohammadi}}, \bibinfo {author}
  {\bibfnamefont {R.~B.}\ \bibnamefont {Staszewski}}, \bibinfo {author}
  {\bibfnamefont {A.}~\bibnamefont {Vladimirescu}}, \bibinfo {author}
  {\bibfnamefont {M.}~\bibnamefont {Babaie}}, \bibinfo {author} {\bibfnamefont
  {F.}~\bibnamefont {Sebastiano}}, \ and\ \bibinfo {author} {\bibfnamefont
  {E.}~\bibnamefont {Charbon}},\ }\bibfield  {title} {\enquote {\bibinfo
  {title} {Cryo-cmos circuits and systems for quantum computing
  applications},}\ }\href {\doibase 10.1109/JSSC.2017.2737549} {\bibfield
  {journal} {\bibinfo  {journal} {IEEE Journal of Solid-State Circuits}\
  }\textbf {\bibinfo {volume} {53}},\ \bibinfo {pages} {309--321} (\bibinfo
  {year} {2018})}\BibitemShut {NoStop}%
\bibitem [{\citenamefont {Valenzuela}\ \emph {et~al.}(2006)\citenamefont
  {Valenzuela}, \citenamefont {Oliver}, \citenamefont {Berns}, \citenamefont
  {Berggren}, \citenamefont {Levitov},\ and\ \citenamefont
  {Orlando}}]{Valenzuela:2006}%
  \BibitemOpen
  \bibfield  {author} {\bibinfo {author} {\bibfnamefont {Sergio~O.}\
  \bibnamefont {Valenzuela}}, \bibinfo {author} {\bibfnamefont {William~D.}\
  \bibnamefont {Oliver}}, \bibinfo {author} {\bibfnamefont {David~M.}\
  \bibnamefont {Berns}}, \bibinfo {author} {\bibfnamefont {Karl~K.}\
  \bibnamefont {Berggren}}, \bibinfo {author} {\bibfnamefont {Leonid~S.}\
  \bibnamefont {Levitov}}, \ and\ \bibinfo {author} {\bibfnamefont {Terry~P.}\
  \bibnamefont {Orlando}},\ }\bibfield  {title} {\enquote {\bibinfo {title}
  {Microwave-induced cooling of a superconducting qubit},}\ }\href {\doibase
  10.1126/science.1134008} {\bibfield  {journal} {\bibinfo  {journal}
  {Science}\ }\textbf {\bibinfo {volume} {314}},\ \bibinfo {pages} {1589--1592}
  (\bibinfo {year} {2006})},\ \Eprint
  {http://arxiv.org/abs/http://science.sciencemag.org/content/314/5805/1589.full.pdf}
  {http://science.sciencemag.org/content/314/5805/1589.full.pdf} \BibitemShut
  {NoStop}%
\bibitem [{\citenamefont {Park}\ \emph {et~al.}(2016)\citenamefont {Park},
  \citenamefont {Rodriguez-Briones}, \citenamefont {Feng}, \citenamefont
  {Rahimi}, \citenamefont {Baugh},\ and\ \citenamefont {Laflamme}}]{Park:2016}%
  \BibitemOpen
  \bibfield  {author} {\bibinfo {author} {\bibfnamefont {Daniel~K.}\
  \bibnamefont {Park}}, \bibinfo {author} {\bibfnamefont {Nayeli~A.}\
  \bibnamefont {Rodriguez-Briones}}, \bibinfo {author} {\bibfnamefont {Guanru}\
  \bibnamefont {Feng}}, \bibinfo {author} {\bibfnamefont {Robabeh}\
  \bibnamefont {Rahimi}}, \bibinfo {author} {\bibfnamefont {Jonathan}\
  \bibnamefont {Baugh}}, \ and\ \bibinfo {author} {\bibfnamefont {Raymond}\
  \bibnamefont {Laflamme}},\ }\enquote {\bibinfo {title} {Heat bath algorithmic
  cooling with spins: Review and prospects},}\ in\ \href {\doibase
  10.1007/978-1-4939-3658-8_8} {\emph {\bibinfo {booktitle} {Electron Spin
  Resonance (ESR) Based Quantum Computing}}},\ \bibinfo {editor} {edited by\
  \bibinfo {editor} {\bibfnamefont {Takeji}\ \bibnamefont {Takui}}, \bibinfo
  {editor} {\bibfnamefont {Lawrence}\ \bibnamefont {Berliner}}, \ and\ \bibinfo
  {editor} {\bibfnamefont {Graeme}\ \bibnamefont {Hanson}}}\ (\bibinfo
  {publisher} {Springer New York},\ \bibinfo {address} {New York, NY},\
  \bibinfo {year} {2016})\ pp.\ \bibinfo {pages} {227--255}\BibitemShut
  {NoStop}%
\bibitem [{Note1()}]{Note1}%
  \BibitemOpen
  \bibinfo {note} {Note that, pin-chip bonding is not limited to this specific
  qubit footprint. Smaller qubit footprints and different qubit designs can me
  considered as well.}\BibitemShut {Stop}%
\bibitem [{Note2()}]{Note2}%
  \BibitemOpen
  \bibinfo {note} {See, e.g., \protect \url
  {http://www.scanditron.com/sites/default/files/material/heraeus_bondingwire_brochure.pdf}
  or \protect \url {https://www.cirexx.com/wire-bonding/}.}\BibitemShut {Stop}%
\bibitem [{\citenamefont {Higginbottom}\ \emph {et~al.}(2018)\citenamefont
  {Higginbottom}, \citenamefont {Campbell}, \citenamefont {Araneda},
  \citenamefont {Fang}, \citenamefont {Colombe}, \citenamefont {Buchler},\ and\
  \citenamefont {Lam}}]{Higginbottom:2018}%
  \BibitemOpen
  \bibfield  {author} {\bibinfo {author} {\bibfnamefont {Daniel~B.}\
  \bibnamefont {Higginbottom}}, \bibinfo {author} {\bibfnamefont {Geoff~T.}\
  \bibnamefont {Campbell}}, \bibinfo {author} {\bibfnamefont {Gabriel}\
  \bibnamefont {Araneda}}, \bibinfo {author} {\bibfnamefont {Fengzhou}\
  \bibnamefont {Fang}}, \bibinfo {author} {\bibfnamefont {Yves}\ \bibnamefont
  {Colombe}}, \bibinfo {author} {\bibfnamefont {Ben~C.}\ \bibnamefont
  {Buchler}}, \ and\ \bibinfo {author} {\bibfnamefont {Ping~Koy}\ \bibnamefont
  {Lam}},\ }\bibfield  {title} {\enquote {\bibinfo {title} {Fabrication of
  ultrahigh-precision hemispherical mirrors for quantum-optics applications},}\
  }\href {\doibase 10.1038/s41598-017-18637-8} {\bibfield  {journal} {\bibinfo
  {journal} {Scientific Reports}\ }\textbf {\bibinfo {volume} {8}} (\bibinfo
  {year} {2018}),\ 10.1038/s41598-017-18637-8}\BibitemShut {NoStop}%
\bibitem [{Note3()}]{Note3}%
  \BibitemOpen
  \bibinfo {note} {See, e.g., \protect \url
  {http://research.physics.illinois.edu/bezryadin/labprotocol/stycast1266.pdf}.}\BibitemShut
  {Stop}%
\bibitem [{\citenamefont {Collin}(2001)}]{Collin:2001}%
  \BibitemOpen
  \bibfield  {author} {\bibinfo {author} {\bibfnamefont {Robert~E.}\
  \bibnamefont {Collin}},\ }\href {\doibase 10.1109/9780470544662} {\emph
  {\bibinfo {title} {Foundations for Microwave Engineering - 2nd Edition}}}\
  (\bibinfo  {publisher} {Institute of Electrical {\&} Electronics Engineers
  ({IEEE}), Inc., and John Wiley {\&} Sons, Inc.},\ \bibinfo {address} {New
  York, {NY}, and Hoboken, {NJ}, {USA}},\ \bibinfo {year} {2001})\BibitemShut
  {NoStop}%
\bibitem [{\citenamefont {Brecht}\ \emph {et~al.}(2017)\citenamefont {Brecht},
  \citenamefont {Chu}, \citenamefont {Axline}, \citenamefont {Pfaff},
  \citenamefont {Blumoff}, \citenamefont {Chou}, \citenamefont {Krayzman},
  \citenamefont {Frunzio},\ and\ \citenamefont {Schoelkopf}}]{Brecht:2017}%
  \BibitemOpen
  \bibfield  {author} {\bibinfo {author} {\bibfnamefont {Teresa}\ \bibnamefont
  {Brecht}}, \bibinfo {author} {\bibfnamefont {Yiwen}\ \bibnamefont {Chu}},
  \bibinfo {author} {\bibfnamefont {Christopher}\ \bibnamefont {Axline}},
  \bibinfo {author} {\bibfnamefont {Wolfgang}\ \bibnamefont {Pfaff}}, \bibinfo
  {author} {\bibfnamefont {Jacob~Z.}\ \bibnamefont {Blumoff}}, \bibinfo
  {author} {\bibfnamefont {Kevin}\ \bibnamefont {Chou}}, \bibinfo {author}
  {\bibfnamefont {Lev}\ \bibnamefont {Krayzman}}, \bibinfo {author}
  {\bibfnamefont {Luigi}\ \bibnamefont {Frunzio}}, \ and\ \bibinfo {author}
  {\bibfnamefont {Robert~J.}\ \bibnamefont {Schoelkopf}},\ }\bibfield  {title}
  {\enquote {\bibinfo {title} {Micromachined integrated quantum circuit
  containing a superconducting qubit},}\ }\href {\doibase
  10.1103/PhysRevApplied.7.044018} {\bibfield  {journal} {\bibinfo  {journal}
  {Phys. Rev. Applied}\ }\textbf {\bibinfo {volume} {7}},\ \bibinfo {pages}
  {044018} (\bibinfo {year} {2017})}\BibitemShut {NoStop}%
\bibitem [{\citenamefont {McRae}\ \emph {et~al.}(2018)\citenamefont {McRae},
  \citenamefont {B\'ejanin}, \citenamefont {Earnest}, \citenamefont {McConkey},
  \citenamefont {Rinehart}, \citenamefont {Deimert}, \citenamefont {Thomas},
  \citenamefont {Wasilewski},\ and\ \citenamefont {Mariantoni}}]{McRae:2018}%
  \BibitemOpen
  \bibfield  {author} {\bibinfo {author} {\bibfnamefont {C.~R.~H.}\
  \bibnamefont {McRae}}, \bibinfo {author} {\bibfnamefont {J.~H.}\ \bibnamefont
  {B\'ejanin}}, \bibinfo {author} {\bibfnamefont {C.~T.}\ \bibnamefont
  {Earnest}}, \bibinfo {author} {\bibfnamefont {T.~G.}\ \bibnamefont
  {McConkey}}, \bibinfo {author} {\bibfnamefont {J.~R.}\ \bibnamefont
  {Rinehart}}, \bibinfo {author} {\bibfnamefont {C.}~\bibnamefont {Deimert}},
  \bibinfo {author} {\bibfnamefont {J.~P.}\ \bibnamefont {Thomas}}, \bibinfo
  {author} {\bibfnamefont {Z.~R.}\ \bibnamefont {Wasilewski}}, \ and\ \bibinfo
  {author} {\bibfnamefont {M.}~\bibnamefont {Mariantoni}},\ }\bibfield  {title}
  {\enquote {\bibinfo {title} {Thin film metrology and microwave loss
  characterization of indium and aluminum/indium superconducting planar
  resonators},}\ }\href {\doibase 10.1063/1.5020514} {\bibfield  {journal}
  {\bibinfo  {journal} {Journal of Applied Physics}\ }\textbf {\bibinfo
  {volume} {123}},\ \bibinfo {pages} {205304} (\bibinfo {year}
  {2018})}\BibitemShut {NoStop}%
\bibitem [{\citenamefont {Fowler}\ \emph {et~al.}(2012)\citenamefont {Fowler},
  \citenamefont {Mariantoni}, \citenamefont {Martinis},\ and\ \citenamefont
  {Cleland}}]{Fowler:2012}%
  \BibitemOpen
  \bibfield  {author} {\bibinfo {author} {\bibfnamefont {Austin~G.}\
  \bibnamefont {Fowler}}, \bibinfo {author} {\bibfnamefont {Matteo}\
  \bibnamefont {Mariantoni}}, \bibinfo {author} {\bibfnamefont {John~M.}\
  \bibnamefont {Martinis}}, \ and\ \bibinfo {author} {\bibfnamefont
  {Andrew~N.}\ \bibnamefont {Cleland}},\ }\bibfield  {title} {\enquote
  {\bibinfo {title} {Surface codes: Towards practical large-scale quantum
  computation},}\ }\href {\doibase 10.1103/PhysRevA.86.032324} {\bibfield
  {journal} {\bibinfo  {journal} {Phys. Rev. A}\ }\textbf {\bibinfo {volume}
  {86}},\ \bibinfo {pages} {032324} (\bibinfo {year} {2012})}\BibitemShut
  {NoStop}%
\end{thebibliography}

%

\end{document}